\documentclass[aps,prc,floatfix,showpacs,twocolumn]{revtex4-1}
\usepackage{ulem}
\usepackage{bm}
\usepackage{color}
\usepackage{amssymb}
\usepackage{amsmath}
\usepackage{graphicx}
\usepackage{amsfonts}
\usepackage{slashed}
\usepackage{pstricks}
\usepackage{float}
\usepackage{hyperref}
\usepackage{array}
\allowdisplaybreaks
\usepackage{dcolumn}
\usepackage{epsf}
\usepackage{tabularx}
\usepackage{soul}
\usepackage{wrapfig}

\begin{document}
\title{Benchmark calculations of infinite neutron matter with realistic two- and three-nucleon potentials}
\author{
A.\ Lovato $^{\, {\rm a,b,c} }$, 
I.\ Bombaci$^{\, {\rm d,e} }$, 
D.\ Logoteta$^{\, {\rm d,e} }$,
M.\ Piarulli$^{\, {\rm f, g} }$,
R.\ B. \ Wiringa$^{\, {\rm a} }$,
}
\affiliation{
$^{\,{\rm a}}$\mbox{Physics Division, Argonne National Laboratory, Argonne, IL 60439, USA}\\
$^{\,{\rm b}}$\mbox{Computational Physics Division, Argonne National Laboratory, Argonne, IL 60439, US}
$^{\,{\rm c}}$\mbox{INFN-TIFPA Trento Institute of Fundamental Physics and Applications, 38123 Trento, Italy}\\
$^{\,{\rm d}}$\mbox{Dipartimento di Fisica ``E. Fermi'', Universit\`a di Pisa, Largo B. Pontecorvo 3, I-56127 Pisa, Italy}\\
$^{\,{\rm e}}$\mbox{INFN, Sezione di Pisa, Largo B. Pontecorvo 3, I-56127 Pisa, Italy}\\
$^{\,{\rm f}}$\mbox{Physics Department, Washington University, St Louis, MO 63130, USA}\\
$^{\,{\rm g}}$\mbox{McDonnell Center for the Space Sciences at Washington University in St. Louis, MO 63130, USA}\\
}
\date{\today}

\begin{abstract}
We present the equation of state of infinite neutron matter as obtained from highly-realistic Hamiltonians that include nucleon-nucleon and three-nucleon  coordinate-space potentials. 
We benchmark three independent many-body methods: Brueckner--Bethe--Goldstone (BBG), Fermi hypernetted chain/single-operator chain (FHNC/SOC), 
and auxiliary-field diffusion Monte Carlo (AFDMC). We find them to provide similar equations of state when the Argonne $v_{18}$ and the Argonne $v_{6}^\prime$ nucleon-nucleon potentials are used in combination with the Urbana IX three-body force. Only at densities larger than about 1.5 the nuclear saturation density 
($\rho_0 = 0.16\,\rm{fm}^{-3}$) the FHNC/SOC energies are appreciably lower than the other two approaches. The AFDMC calculations carried out with all of the Norfolk potentials fitted to reproduce the experimental trinucleon ground-state energies and $nd$ doublet scattering length yield unphysically bound neutron matter, associated with the formation of neutron droplets. Including tritium $\beta$-decay in the fitting procedure, as in the second family of Norfolk potentials, mitigates but does not completely resolve this problem. An excellent agreement between the BBG and AFDMC results is found for the subset of Norfolk interactions that do not make neutron-matter collapse, while the FHNC/SOC equations of state are moderately softer. 
\end{abstract}
\maketitle


\section{Introduction} 
Connecting properties of atomic nuclei to the equation of state (EOS) of infinite nucleonic matter within a microscopic perspective, in which nuclear systems are described in terms of nucleon-nucleon ($N\!N$) and three-nucleon ($3N$) forces, as well as consistent electroweak currents, is a long-standing challenge for nuclear physics. Addressing it has become extremely timely as neutron stars, which contain the universe's most dense nuclear materials, can now be probed in whole new ways from gravitational waves to satellite X-ray telescopes, providing an opportunity to test the high-density and low temperature regime of matter that 
is not currently accessible by terrestrial experiments~\cite{Sabatucci:2020xwt,senger2021}. 
Indeed, the first observation of gravitational waves, GW170817, from a binary neutron star merger~\cite{LIGOScientific:2017vwq}, in conjunction with its electromagnetic counterpart~\cite{Abbott:2017}, has ushered in a new era in our quest to address some of the most fundamental questions driving nuclear physics today. The gravitational wave signal emitted during binary neutron star inspirals is strongly influenced by the tidal deformability of the participating stars, which is in turn largely determined by the nuclear matter EOS~\cite{Hinderer:2009ca}. 
Using GW170817 data, the LIGO-Virgo Collaboration has been able to constrain the value of the adimensional tidal deformability $\Lambda_{1.4}$, for a neutron star with a mass $M = 1.4\, M_\odot$, to be in the range  \cite{LIGOScientific:2018cki}  $\Lambda_{1.4} = 190^{+390}_{-120}$. 
This implies small stellar radii in the range $R_{1.4} = 11.9 \pm 1.4\,\rm{km}$ which points to a soft EOS at densities of about 2-3 times the nuclear matter saturation density $\rho_0 = 0.16\, \rm{fm}^{-3}$.  

Additional observational constraints on the nuclear EOS comes from measured neutron star masses. 
Particularly three neutron stars with masses close to $M =2\,M_\odot$ have been reliably established over the last decade. 
Indeed, pulsar-timing observations of the millisecond pulsar J0740+6620 have yielded the most massive neutron star 
yet observed: $M=2.14^{+0.10}_{-0.09} \,M_\odot$~\cite{NANOGrav:2019jur}. 
The existence of neutron stars with masses $M \sim 2\, M_\odot$ distinctly indicates a stiff EOS at $\rho \gtrsim 3 \rho_0$ to support them. There are several EOS models 
(e.g. WFF \cite{WFF88}, APR \cite{Akmal:1998cf}, BL \cite{BL2018}) that can reconcile a soft EOS at $\rho \sim (2$--$3)\, \rho_0$, as suggested by GW170817, with a stiff EOS at  $\rho \gtrsim 3 \rho_0$ to support massive neutron stars against gravitational collapse to black holes.  

The recent simultaneous determination of the mass and radius of a few neutron stars by the Neutron star Interior Composition Explorer (NICER) aboard the International Space Station is providing new and significant information to constrain the nuclear EOS. In the case of the neutron star J0030+0451 the NICER collaboration applied different analyses on the observational data and obtained values of (all at 68\% credibility) $R=12.71^{+1.14}_{-1.19}$\,km and $M=1.34^{+0.15}_{-0.16} \,M_\odot$~\cite{Riley:2019yda} and $R=13.02^{+1.24}_{-1.06}$\,km and $M=1.44^{+0.15}_{-0.14} \,M_\odot$~\cite{Miller:2019cac}, which are consistent with the constraints on the stellar radius derived from GW170817  data. Another recent measurement by the NICER team was of the pulsar PSR J0740+6620, which is of major interest to nuclear physicists due to its much heavier mass, $M=2.08^{+0.07}_{-0.07} \,M_\odot$~\cite{Fonseca:2021wxt} (this is an update from the original value reported by~\cite{NANOGrav:2019jur}). Two independent analyses found the radius to be $R=12.39^{+1.3}_{-1.06}$\,km~\cite{Riley:2021pdl} 
and $R=13.71^{+2.61}_{-1.50}$\,km ~\cite{Miller:2021qha} at 68\% credibility.

In addition to the above astrophysics implications, precise calculations of pure neutron matter (PNM) EOS have become relevant for the construction of nuclear energy density functionals (EDFs). Traditional EDFs are able to reproduce a myriad of nuclear observables but are biased towards experimentally well-known regions, e.g. nuclei close to the stability valley~\cite{Brown:2013pwa}. Novel strategies to constrain the EDF on microscopic calculations of PNM have been developed~\cite{Brown:2013pwa,Bulgac:2017bho,Tsang:2019ymt,Buraczynski:2019yom,Marino:2021xyd,Yang:2021akb} to accompany the growing experimental research efforts devoted to unstable nuclei at the limits of the nuclear chart. Their success however, relies on accurate microscopic calculations of the EOS, supplemented by reliable estimates of theoretical uncertainties. 

Despite recent remarkable progress, it is still unclear whether there exist microscopic nuclear potentials that accurately describe light nuclear systems --- e.g. nucleon-nucleon scattering, light nuclear spectra, densities, structure functions, transitions, and responses --- and the EOS of infinite neutron matter. For example, the phenomenological Argonne $v_{18}$~\cite{Wiringa:1994wb} plus Illinois 7~\cite{Pieper:2008rui} (AV18+IL7) Hamiltonian can reproduce the experimental energies of nuclei with up to $A=12$ with high precision~\cite{Carlson:2014vla}, but it fails to provide sufficient repulsion in pure neutron matter~\cite{Maris:2013rgq}. On the other hand, the Argonne $v_{18}$ plus Urbana IX~\cite{Pudliner:1995wk} model (AV18+UIX), which provides a good description of nuclear-matter properties and the basis for the highly-cited APR beta-stable EOS for neutron stars~\cite{Akmal:1998cf}, does not give as satisfactory a reproduction of the spectrum of light nuclei~\cite{Pieper:2001ap}. These interactions being largely phenomenological, no clear prescriptions are currently available to properly assess their uncertainties and to systematically improve them, especially as far as three-nucleon forces are concerned~\cite{Piarulli:2019cqu}. Over the last decades, chiral effective field theory ($\chi$EFT) has been extensively employed to systematically derive high-quality $N\!N$ and $3N$ interactions~\cite{Epelbaum:2008ga,Machleidt:2011zz,Reinert:2017usi,Entem:2017gor}. Local, coordinate-space versions of $\chi$EFT $3N$ forces were first included in quantum Monte Carlo methods in Ref.~\cite{Lovato:2011ij}, in combination with phenomenological Argonne $v_6^\prime$ (AV6P) $N\!N$ potential. Local, $\chi$EFT $N\!N$ potentials were first implemented in quantum Monte Carlo methods by the authors of Ref.~\cite{Gezerlis:2013ipa}. As a major breakthrough, the results presented in Ref.~\cite{Gezerlis:2013ipa, Lynn:2015jua} also include consistent coordinate-space $\chi$EFT $3N$ forces that were fit to reproduce the binding energy of $^4$He and low-energy neutron - $^4$He scattering. These interactions have been subsequently employed to carry out quantum Monte Carlo calculation of the equation of state of symmetric nuclear matter~\cite{Lonardoni:2019ypg}, and have proven to reproduce empirical saturation density and energy well within statistical and systematic uncertainties. It has to be noted that even state-of-the-art $\chi$EFT interactions do not fully solve the tension between light systems and infinite matter. For instance, the potentials developed in Ref.~\cite{Jiang:2020the} yield binding energies and radii for a wide range of nuclei that are compatible with experiments together with an accurate nuclear matter EOS, but fail to reproduce proton-proton scattering data below $100$ MeV laboratory energy~\cite{Nosyk:2021pxb}.

In this work, we employ the first and second generation of $\chi$EFT Norfolk $N\!N$ and $3N$ interactions that provide accurate description of the spectra and low-energy transitions of light nuclei~\cite{Piarulli:2017dwd,Baroni:2018fdn}. These two generations differ in the strategy adopted to determine the low-energy constants entering the $3N$ force. The first one uses the experimental trinucleon ground-state energies and $nd$ doublet scattering length, while the second includes in the fitting procedure the empirical value of the Gamow-Teller matrix element of tritium $\beta$ decay. We test their predictions in neutron matter using three many-body approaches: the Brueckner--Bethe--Goldstone (BBG)~\cite{bbg1,baldo-1999}, the Fermi hypernetted chain/single-operator chain (FHNC/SOC)~\cite{FR75,PW79}, and the auxiliary-field diffusion Monte Carlo (AFDMC)~\cite{Schmidt:1999lik}. In our analysis we also consider the widely-used AV18+UIX and its simplification AV6P+UIX.
In the same spirit as our previous work~\cite{Piarulli:2019pfq}, which only included $N\!N$ potentials, by performing these benchmark calculations we aim at better assessing the systematic error associated with the method used for solving the many-body Schr\"odinger equation and quantifying its relative importance compared to the uncertainties inherent to the nuclear Hamiltonian. While benchmarks of the various many-body methods have been regularly undertaken for nuclei~\cite{Kamada:2001tv,Hagen:2007ew,Tichai:2020dna}, with the exception of Ref.~\cite{Baldo:2012nh,Piarulli:2019pfq}, they have not been as frequently carried out for PNM.  

This work is organized as follows. In Section~\ref{sec:nuc_ham}, we introduce the nuclear Hamiltonians with particular emphasis on the $3N$ force. The many-body methods employed for calculating the EOS of PNM are summarized in Section~\ref{sec:mbm}. The benchmark calculations obtained within three different many-body methods are discussed in Section~\ref{sec:results}. Finally, in Section~\ref{sec:conclusions} we draw our conclusions and provide some perspective on future research directions. 

\section{Nuclear Hamiltonians}
\label{sec:nuc_ham}
Our PNM calculations are based on the well-known AV18+UIX~\cite{Wiringa:1994wb,Pudliner:1995wk} model as well as the Norfolk $N\!N$ and $3N$ interactions developed in Ref.~\cite{Piarulli:2016vel,Piarulli:2014bda,Baroni:2018fdn}. The latter are referred to in the literature as NV2+3. The NV2+3 potentials are formulated in coordinate space and derived from chiral effective field theory ($\chi$EFT) in which pions, nucleons and $\Delta$'s are retained as fundamental degrees of freedom. They include long-range components mediated by one- and two-pion exchange, as well as contact terms characterized by unknown low-energy constants (LECs). Specifically, the two-body component (NV2) is constructed up to N3LO in the chiral expansion, retaining only contact terms at this order, and the three-body force (NV3) includes corrections up to N2LO. 

The LECs that enter the NV2 contact interactions are constrained to reproduce $N\!N$ scattering data from the most recent database collected by the Granada group~\cite{Perez:2013jpa,Perez:2013oba}. The contact terms are regularized via a Gaussian cutoff function
with $R_S$ as the Gaussian parameter~\cite{Piarulli:2016vel,Piarulli:2014bda,Baroni:2018fdn},
\begin{equation}
C_{R_{\rm S}}(r) = \frac{1}{\pi^{3/2}\, R_{\rm S}^3} {\rm e}^{-(r/R_{\rm S})^2} \ ,
\end{equation}
while the divergences appearing at high values of momentum transfer in the pion-range operators are removed via a special radial function characterized by the cutoff $R_L$~\cite{Piarulli:2016vel,Piarulli:2014bda,Baroni:2018fdn},
\begin{equation}
C_{R_{\rm L}}(r)= 1-\frac{1}{(r/R_{\rm L})^6 \, {\rm e}^{(r-R_{\rm L})/a_{\rm L}}+1} \ ,
\end{equation}
with the diffuseness $a_L$ being fixed at $a_L = R_L/2$. There are two classes of NV2 potentials. Class I (II) has been fitted to $N\!N$ data up to $125$ MeV ($200$ MeV). For each class, two combinations of short- and long-range regulators have been used, namely ($R_S$, $R_L$)=(0.8, 1.2) fm (models NV2-Ia and NV2-IIa) and 
($R_S$, $R_L$)=(0.7, 1.0) fm (models NV2-Ib and NV2-IIb). Class I (II) fits about 2700 (3700) data points with a $\chi^2$/datum $\lesssim 1.1$ 
($\lesssim 1.4$)~\cite{Piarulli:2016vel,Piarulli:2014bda}. 

The NV3 $3N$ potentials~\cite{Piarulli:2017dwd} are the sum of three contributions
\begin{equation}
V=\sum_{i<j<k}\sum_{{\rm cyc}} \left[V^{\Delta}(ijk)+ V^{2\pi}(ijk)+V^{\rm CT}(ijk)\right] \ ,
\end{equation}
where ``cyc'' denotes the cyclic permutations of particles $i$, $j$, and $k$. The terms $V^{\Delta}(ijk)$ and $V^{2\pi}(ijk)$ represent the long-range piece of the $3N$ force mediated by the two-pion exchange (TPE) diagrams with and without $\Delta$'s, respectively, and $V^{\rm CT}(ijk)$ is the short-range piece, which is parametrized by contact terms. 

Considering that the isospin operator ${\bm \tau}_i\cdot {\bm \tau}_j=1$ in PNM, the expressions for $V^{\Delta}(ijk)$,  $V^{2\pi}(ijk)$, and $V^{\rm CT}(ijk)$ relevant for this work are:
\begin{eqnarray}
\label{eq:e51_delta}
V^\Delta(ijk) &=&  -  \frac{g_A^2 h_A^2} 
{ 72\cdot 144\, \pi^2} \,\frac{m_\pi^6} {m_{\Delta N}f_\pi^4}
2\,\Big \{ \widetilde{X}_{ij}\, ,\, \widetilde{X}_{jk} \Big\} \\
\label{eq:e51_2pi}
V^{2\pi}(ijk)&=&\frac{g_A^2}{256\, \pi^2} \frac{m_\pi^6}{f_\pi^4}\bigg[
-16\, c_1\, \widetilde{Z}_\pi(r_{ij}) \, \widetilde{Z}_\pi(r_{jk})\nonumber\\
&&\times{\bm \sigma}_i\cdot\hat{\bf r}_{ij}
\,\,{\bm \sigma}_k\cdot\hat{\bf r}_{jk}+\frac{4}{9}\, c_3 
\left\{  \widetilde{X}_{ij} \, , \, \widetilde{X}_{jk} \right\}\bigg ]\\
\label{eq:e51_ct}
 V^{\rm CT}(ijk)&=& \frac{g_A\, c_D}{96\, \pi} \, \frac{m_\pi^3}{\Lambda_\chi\, f_\pi^4}\,
\widetilde{X}_{ik}\,  \left[ C_{R_{\rm S}}(r_{ij}) +C_{R_{\rm S}}(r_{jk}) \right] \nonumber\\
&& +\frac{c_E}{\Lambda_\chi\, f_\pi^4}\, C_{R_{\rm S}}(r_{ij})\, C_{R_{\rm S}}(r_{jk})\ ,
\end{eqnarray}
In the above equations, $g_A$ and $h_A$ are the axial and the $N$-to-$\Delta$ axial coupling constants, respectively, $m_{\pi}$ and $m_{\Delta N}$ are the pion and $\Delta-N$ mass difference, respectively, $f_{\pi}$ is the decay amplitude, and $\Lambda_{\chi} =1$ GeV is the chiral-symmetry-breaking scale. Their values are reported in Table I and II of Ref.~\cite{Piarulli:2014bda}. The operator $\widetilde{X}_{ij}$ is defined as 
\begin{equation}
\widetilde{X}_{ij} = \widetilde{T}_{\pi}(r_{ij})\, S_{ij} + \widetilde{Y}_{\pi}(r_{ij})\,
{\bm \sigma}_i \cdot {\bm \sigma}_j \, ,
\end{equation}
where the regularized functions are
\begin{eqnarray}
\widetilde{Y}_\pi(r)&=&\frac{e^{-{m_{\pi}}r}}{{m_{\pi}}r} \,C_{R_{\rm L}}(r) \ , \\
\widetilde{T}_{\pi}(r)&=&\left( 1 + \frac{3}{m_{\pi}\,r} + \frac{3}{m_{\pi}^2\, r^2} \right)  \widetilde{Y}_{\pi}(r) \ , \\
\widetilde{Z}_\pi(r)&=&-\left( 1+\frac{1}{m_\pi\, r}\right)\widetilde{Y}_\pi(r) \ .
\end{eqnarray}

The TPE contribution in Eq.~(\ref{eq:e51_2pi}) depends on the pion-nucleon LECs $c_1$, $c_3$, and $c_4$, that already appear in the $N\!N$ sector, and their values are listed in Table I of Ref.~\cite{Piarulli:2014bda}. The short-range component, instead, is parametrized in terms of two unknown LECs, $c_D$ and $c_E$. In the first generation of Norfolk potentials
(NV2+3-Ia/b and NV2+3-IIa/b), these LECs have been determined by simultaneously reproducing the experimental 
trinucleon ground-state energies and $nd$ doublet scattering length~\cite{Piarulli:2017dwd}. Within the $\chi$EFT framework, $c_D$ is related to the LEC entering the axial two-body contact current~\cite{Gazit:2008ma,Marcucci:2011jm,Schiavilla:2017}, allowing one to adopt a different strategy to constrain $c_D$ and $c_E$.
In particular, in Ref.~\cite{Baroni:2018fdn}, they have been determined
to reproduce the trinucleon binding energies and the empirical value of the Gamow-Teller matrix element in tritium $\beta$ decay. Norfolk models that use this fitting procedure are designated with a `*' namely, NV2+3-Ia*/b* and NV2+3-IIa*/b*. For completeness, we report the values of $c_D$ and $c_E$ in Table~\ref{tab:cdce}.

\begin{center}
	\begin{table}
	\resizebox{245pt}{!}{
		\begin{tabular}{l|r r r r}
			\hline\hline
			       & Ia  (Ia*)         &        Ib (Ib*)   &      IIa (IIa*) & IIb (IIb*)\\
			\hline
			$c_D$  &    3.666 (--0.635)  &  --2.061 (--4.71) &   1.278 (--0.61)    & --4.480 (--5.25) \\
			$c_E$  &  --1.638 (--0.090)  &  --0.982 (  0.55) & --1.029 (--0.35)    & --0.412 (0.05)\\
				\hline\hline
		\end{tabular}
		}
		\caption{Adimensional $c_D$ and $c_E$ values of the contact terms in the NV3
			interactions obtained from fits to  {\it i}) the $nd$ scattering length and 
			trinucleon binding energies~\cite{Piarulli:2017dwd}; and {\it ii}) the 
			central value of the $^3$H GT matrix element and the trinucleon binding energies (starred
			values)~\cite{Baroni:2018fdn}.
				}
		\label{tab:cdce}
	\end{table}
\end{center}

\section{Many-body methods}
\label{sec:mbm}

\subsection{BBG}
The Brueckner--Bethe--Goldstone (BBG) many-body theory (see e.g., \cite{bbg1,baldo-1999}) is based on a 
linked cluster expansion (the so-called hole-line expansion) of the energy per nucleon $E/A$ of nuclear matter.   
The various terms of the expansion can be represented by Goldstone diagrams \cite{goldstone57} 
grouped according to the number of independent hole-lines (i.e., lines representing empty single 
particle states in the Fermi sea). 
The basic ingredient in this approach is the Brueckner reaction matrix $G$ \cite{brueck1,brueck2} 
which sums, in a closed form the infinite series of the so-called ladder-diagrams and allows treatment of 
the short-range strongly repulsive part of the nucleon-nucleon interaction.   
The G-matrix can be obtained by solving the Bethe--Goldstone equation \cite{BG57}  
\begin{equation}
 G(\omega) =  V  + V  \sum_{k_a,k_b} 
\frac{\mid{\bf k_a},{\bf k_b}\rangle \, Q\, \langle{\bf k_a},{\bf k_b}\mid}
     {\omega - \epsilon(k_a) - \epsilon(k_b) + i\eta }\, G(\omega) \;,
\label{bg} 
\end{equation}
where $V$ is the bare $N\!N$ interaction (or a density dependent two-body effective interaction when three-nucleon forces are introduced 
as discussed below) the quantity $\omega$ is the so-called starting energy. 
In the present work we consider spin-unpolarized neutron matter, thus in equation (\ref{bg}) and in the 
following equations we drop the spin indices to simplify the mathematical notation.    
The Pauli operator $\mid{\bf k_a},{\bf k_b}\rangle Q \langle {\bf k_a},{\bf k_b}\mid$ 
projects on intermediate scattering states in which the momenta ${\bf k_a}$ and ${\bf k_b}$
of the two interacting neutrons are above their Fermi momentum $k_{F}$ since single particle states 
with momenta smaller that this value are occupied by the neutrons of the nuclear medium.   
Thus the Bethe--Goldstone equation describes the scattering of two nucleons (two neutrons in our case) 
in the presence of other nucleons, and the Brueckner $G$-matrix represents the effective interaction between 
two nucleons in the nuclear medium and properly takes into account the short-range correlations arising from 
the strongly repulsive core in the bare $N\!N$ interaction.  

The single-particle energy $\epsilon(k)$ of a neutron with momentum ${\bf k}$, appearing in the energy 
denominator of the Bethe--Goldstone equation (\ref{bg}), is given by 
\begin{equation}
       \epsilon(k) = \frac{\hbar^2 k^2}{2m} + U(k) \ ,
\label{spe}
\end{equation}
where $U(k)$ is a single-particle potential which represents the mean field felt by a neutron  
due to its interaction with the other neutrons of the medium. 
In the Brueckner--Hartree--Fock (BHF) approximation of the BBG theory, $U(k)$ 
is calculated through the real part of the $G$-matrix \cite{BBP63,HM72} and is given by 
\begin{equation}
 U(k) = \sum_{k^\prime \leq k_{F}} 
            \mbox{Re} \ \langle {\bf k},{\bf k^\prime} \mid 
               G(\omega^*) \mid {\bf k},{\bf k^\prime} \rangle_A  \;,
\label{spp}
\end{equation}
where the sum runs over all neutron occupied states, the starting energy is 
$\omega = \omega^* \equiv \epsilon(k) + \epsilon(k')$ (i.e., the G-matrix is calculated on-the-energy-shell) 
and the matrix elements are properly antisymmetrized. 
We make use of the so-called continuous choice \cite{jeuk+67,gra87,baldo+90,baldo+91} for 
the single-particle potential $U(k)$  when solving the Bethe--Goldstone equation.    
As it has been shown in Ref. \cite{song98,baldo00}, the contribution of the three-hole-line diagrams 
to the energy per nucleon $E/A$ is minimized in this prescription for the single particle potential and 
a faster convergence of the hole-line expansion for $E/A$ is achieved with respect to the so-called 
gap choice for $U(k)$.  

In this scheme Eqs.\ (\ref{bg})--(\ref{spp}) have to be solved self-consistently 
using an iterative numerical procedure. 
Once a self-consistent solution is achieved, the energy per nucleon of the system 
can be evaluated in the BHF approximation of the BBG hole line-expansion 
and it is given by    
\begin{equation}
 \frac{E}{A} = \frac{1}{A} \sum_{k < k_{F}}
    \left(\frac{\hbar^2 k^2}{2m} + \frac{1}{2} U(k) \right) \ .
\label{bea}
\end{equation}

Making the usual angular average of the Pauli operator and of the energy denominator \cite{gra87,baldo+91}, 
the Bethe--Goldstone equation (\ref{bg}) can be expanded in partial waves.  
In all the calculations performed in this work, we have considered partial wave contributions up to a total 
two-body angular momentum $J_{max} = 9$.  
We have verified that the inclusion of partial waves with $J_{max} > 9$ does not appreciably change our results.

Within the BHF approach three-nucleon forces  cannot be used directly in their original form. 
This is because it would be necessary to solve three-body Faddeev equations in the nuclear medium 
(Bethe--Faddeev equations) \cite{bethe65,rajaraman-bethe67} and currently this is a task still far from achievement.  
To circumvent this problem an effective density dependent two-body force $V_{eff}(\rho)$ is built starting from the original three-body one by averaging over one of the three nucleons \cite{loiseau,grange89,LBK-2016}
and this is added to the bare $N\!N$ interaction to solve the Bethe--Goldstone equation (\ref{bg}).  

In the present work, we construct this effective density dependent two-body force as follows: 
we first antitransform the coordinate-space three-nucleon potential to momentum space, then we average over the momentum and spin of one of the nucleons using: 
\begin{equation}\label{eq:normord_singpart}
  V_{eff} =  \text{Tr}_{(\sigma_3)} \int \frac{d \bm{p}_3}{(2 \pi)^3} \, n_{\bm{p}_3} \, V_{123} \, (1-P_{13}-P_{23})\, ,
\end{equation}
where 
\begin{equation}  
  P_{ij} = \frac{1 + \bm\sigma_i\cdot\bm\sigma_j}{2} \, P_{\bm p_i \leftrightarrow\bm p_j} 
\end{equation} 
are spin-momentum exchange operators and $n_{\bm{p}_3}$ is the momentum distribution of the third nucleon. 
Here we do not consider the effects of $N\!N$ correlations on the nucleon momentum distribution \cite{Baldo:1991sce,Baldo:1992vcf,Muther:2000qx} 
and we assume for $n_{\bm{p}_3}$ a step function approximation.  

An important point of the whole BHF calculation is that when the three-nucleon force is added to the bare $N\!N$ interaction, 
some double counting is introduced both in the calculation of single-particle potentials and in the total energy per particle. 
In order to take care of this issue, we adopt the strategy suggested in Refs. \cite{Arellano:2016crt, Carbone:2013rca} namely we subtract from the single-particle potentials $1/2$ 
of the Hartree-Fock contribution due to the contribution of the three-nucleon forces. We finally correct the total energy per particle 
with the appropriate Hartree-Fock correction due to the inclusion of the three-body forces \cite{Arellano:2016crt, Carbone:2013rca}.

\subsection{FHNC/SOC}
In the absence of interactions, a uniform system of $A$ non-interacting neutrons can be described as a Fermi gas at zero temperature, and its ground state wave function reduces to the Slater determinant of orbitals associated with the single-particle states belonging to the Fermi sea
\begin{equation}
\Phi(X)=\mathcal{A}[\,\phi_{n_1}(x_1)\dots\phi_{n_A}(x_A)\,]\, .
\label{def:phi}
\end{equation}
In the above equation $X=\{x_1,\dots,x_A\}$, where the generalized coordinate $x_i\equiv\{\mathbf{r}_i,s_i\}$ represents both the position $R=\mathbf{r}_1,\dots,\mathbf{r}_A$ and the spin $S=s_1,\dots,s_A$, variables of the $i$-th nucleon while $n_i$ denotes the set of quantum numbers specifying the single particle state. 
Translational invariance imposes that the single-particle wave functions be plane waves
\begin{equation}
\phi_{n_i}(x_i)= \frac{1}{\sqrt{\Omega}} {\rm e}^{i\mathbf{k_i}\cdot\mathbf{r}_i} \chi_{\sigma_i}(s_i)
\end{equation}
In the above equations, $\Omega$ is the normalization volume, $\chi_{\sigma_i}(s_i)$ is the spinor of the neutron and $|\mathbf{k}_i|<k_F=(3\pi^2\rho)^{1/3}$. 
Here $k_F$ is the Fermi momentum and $\rho$ the density of the system. 

The variational ansatz of the Fermi hypernetted chain (FHNC) and single-operator chain (SOC) formalism emerges as a generalization of the Jastrow theory of Fermi liquids~\cite{jastrow55,PW79}
\begin{equation}
|\Psi_T\rangle=\frac{F|\Phi\rangle}{\langle\Phi|F^\dagger F|\Phi\rangle^{1/2}}\, .
\label{eq:psi_FHNC}
\end{equation}
where $|\Phi\rangle$ is the Slater determinant of Eq.~\eqref{def:phi} and 
\begin{equation}
F(x_1,\dots,x_A)=\mathcal{S} \left( \prod_{j>i=1}^A F_{ij} \right)
\label{eq:Foperator}
\end{equation}
is the correlation operator. 
The spin-isospin structure of $F_{ij}$ reflects that of the dominant parts of the nucleon-nucleon potential
\begin{equation}
F_{ij} = \sum_{p=1}^8 f^{p}(r_{ij})O^{p}_{ij} \ ,
\label{eq:Foperator_v8prime}
\end{equation}
where the eight operators $O^p_{ij}$ are 1, ${\bm \sigma}_i \cdot {\bm \sigma}_j$, $S_{ij}$, and $\mathbf{L}\cdot\mathbf{S}$, and each of these times ${\bm \tau}_i \cdot {\bm \tau}_j$.  
For pure neutron matter, the expectation of the pair isospin operator is unity, reducing the effective number of $f^p$ components to four.  
Since, in general, $[O^{p}_{ij},O^{q}_{ik}] \neq 0$, the symmetrization operator $\mathcal{S}$ is needed to fulfill the requirement of antisymmetrization of the wave-function.
The $f^{p}(r_{ij})$ are finite-ranged functions, with the conditions
\begin{align}
f^{p}(r\geq d_p) &= \delta_{p1}\, , \nonumber \\
\frac{df^p(r)}{dr}\Big|_{r=d_p} &= 0\, .
\label{eq:heal}
\end{align}
where the $d_p$ are ``healing distances''.  
Consequently, the correlation operator of Eq.~\eqref{eq:Foperator} respects the cluster property: if the system is split in two (or more) subsets of particles that are moved far away from each other, the $F$ factorizes into a product of two factors in such a way that only particles belonging to the same subset are correlated. 

The radial functions $f^{p}(r_{ij})$ are determined by minimizing the energy expectation value
\begin{equation} 
 E_V=\langle\Psi_T|H|\Psi_T\rangle \geq E_0\, , 
\label{eq:ev}
\end{equation}
which provides an upper bound to the true ground state energy $E_0$. 
The energy expectation value in matter is evaluated using a diagrammatic cluster expansion and a set of 29 coupled integral equations, which effectively make partial summations to infinite order -- the FHNC/SOC approximation~\cite{PW79}.  
This is a generalization of the original hypernetted chain (HNC) method for Bose systems developed by van Leeuwen, Groeneveld, and de Boer~\cite{vLGdB59}, which requires the solution of a single integral equation, and the corresponding extension for spin-isospin independent Fermi systems by Fantoni and Rosati~\cite{FR75}, which requires four coupled integral equations.
The integral equations are used to generate two- and three-body distribution functions $g_2(r_{ij}) \equiv g_{ij}$ and $g_3({\bf r}_{ij},{\bf r}_{ik}) \equiv g_{ijk}$, which can then be used to evaluate the energy or other operators.  

For the pure Jastrow case, we evaluate the Pandharipande--Bethe~\cite{PB73} expression for the energy:
\begin{equation}
E_{PB} = T_F + W + W_F + U + U_F \ ,
\label{eq:epb}
\end{equation}
where $T_F$ is the Fermi gas kinetic energy. The only terms for a Bose system are
\begin{align}
  W &= \frac{\rho}{2} \int \Big( v_{ij} - \frac{\hbar^2}{m}
        \frac{\nabla^2 F_{ij}}{F_{ij}} \Big) g_{ij} d^3r_{ij} \ , \nonumber \\
  U &= - \frac{\hbar^2}{2m} \frac{\rho^2}{4} \int
  \Big(\frac{{\bf\nabla}_i F_{ij}\cdot {\bf\nabla}_i F_{ik}}{F_{ij}F_{ik}}\Big)
             g_{ijk} d^3r_{ij} d^3r_{ik}\, ,
\end{align}
while $W_F$, $U_F$ are additional two- and  three-body kinetic energy terms present due to the Slater determinant.
Alternately we use the Jackson--Feenberg~\cite{JF62} energy expression
\begin{eqnarray}
E_{JF} &=& T_F + W_B + W_\phi + U_\phi \ , \\
  W_B &=& \frac{\rho}{2} \int \Big[v_{ij}- \frac{\hbar^2}{2m}
\Big(\frac{\nabla^2 f_{ij}}{f_{ij}} - \frac{(\nabla_i f_{ij})^2}{f^2_{ij}} \Big) \Big] \ ,
\label{eq:ejf}
\end{eqnarray}
where $W_B$ is the boson term and $W_\phi$ and $U_\phi$ are kinetic energy terms involving the Slater determinant.
In principle, these energies should be equivalent, but in practice there are differences due to the FHNC/SOC approximation to the distribution functions.
We take the average $E_V = (E_{PB} + E_{JF})/2$ as our energy expectation value and the difference $\delta E_V = |E_{PB} - E_{JF}|/2$ as an estimate of the error in the calculation.

The FHNC two-body distribution function can be written as:
\begin{eqnarray}
g_{ij} &=& f^2 \Big[ (1+G_{de}+{\mathcal E}_{de})^2 + G_{ee}+{\mathcal E}_{ee} \nonumber \\
       &-& \nu(G_{cc}+{\mathcal E}_{cc}-\ell/\nu)^{2} \Big]
           \exp(G_{dd}+{\mathcal E}_{dd}) \ .
\label{eq:gij}
\end{eqnarray}
where the chain functions $G_{xy}$ are sums of nodal diagrams, with direct ($d$), exchange ($e$) or circular exchange ($c$) end points, ${\mathcal E}_{xy}$ are elementary diagrams, $\ell \equiv \ell(k_Fr)$ is the Slater function, and $\nu$ is the degeneracy.
An example of the structure of the integral equations is:
\begin{align}
G_{dd,ij} &= \rho\int d^{3}r_{k} \left[ ( X_{dd,ik} + X_{de,ik} ) S_{dd,kj} \right. \nonumber\\
&\left. + X_{dd,ij}S_{de,kj} \right] \ ,
\end{align}
where $S_{dd} = f^{2}\exp(G_{dd}+E_{dd}) - 1$ is a two-point superbond and $X_{dd} = S_{dd} - G_{dd}$ is a link function.

The introduction of spin-isospin correlations with operators that do not commute complicates the calculation.  
Fortunately, the first six operators $p=1,6$ form a closed spin-isospin algebra, allowing single continuous chains of operator links -- the SOCs -- to be evaluated.  
These involve five chain functions $G^p_{xy}$ for each of the five operators $p=2-6$, with $xy = dd, de, ee, ca, cb$ in addition to the four Jastrow chain functions in Eq.\eqref{eq:gij}, making the total of 29 coupled integral equations to be solved in nuclear matter, which reduces to 14 coupled integral equations in pure neutron matter.
There are significant contributions from unlinked diagrams in the SOC cluster expansion, but these can be accommodated by means of ``vertex'' corrections, as discussed in Ref.~\cite{PW79}.
Additional higher-order corrections coming from (parallel) multiple operator chains and rings are also calculated, as discussed in Ref.~\cite{WFF88}.

As opposed to the FHNC/SOC calculations reported in Ref.~\cite{Baldo:2012nh}, in this work  we include spin-orbit correlations, corresponding to the $p=7,8$ terms in Eq.~\eqref{eq:Foperator_v8prime}.  
Because of the presence of a derivative operator, these correlations cannot be ``chained'' so they are treated explicitly only at the two- and three-body cluster level. 
It has to be noted that while the two-body cluster contribution is evaluated exactly, following the prescription of Ref.~\cite{WFF88} only a limited number of three-body terms in the cluster expansion are kept.

In standard FHNC calculations, the elementary diagrams of Eq.~\eqref{eq:gij} are generally neglected.  Inclusion of the leading four-body elementary diagram leads to the FHNC/4 approximation~\cite{Zabo77}, while additional contributions have been studied in liquid atomic helium systems~\cite{UP82}.
In the present work we include many central ($p=1$) ${\mathcal E}_{xy}$ diagrams, beyond the FHNC/4 approximation, by introducing three-point superbonds $S_{xyz}$, such as
\begin{eqnarray}
S_{ddd,123} &=& \rho\int d^{3}r_{4} \left\{ S_{dd,14}S_{dd,24}
                   ( S_{dd,34} + S_{de,34} ) \right. \nonumber \\
  &+& ( S_{dd,14}S_{de,24} + S_{de,14}S_{dd,24} ) S_{dd,34} \left.\right\} \ ,
\end{eqnarray}
and then evaluating
\begin{eqnarray}
   {\mathcal E}_{dd,12} &=& \frac{1}{2}\rho\int d^{3}r_{3}
                 \left\{ S_{ddd,132}\left[ S_{dd,13}( S_{dd,32} + S_{de,32} )
                 \right.\right. \nonumber\\
             &+& S_{de,13}S_{dd,32} \left.\right]
               + S_{ded,132}S_{dd,24}S_{dd,32} \left.\right\} \ .
\end{eqnarray}
With six $S_{xyz}$, where $xyz$ = $ddd$, $dde$, $dee$, $eee$, $ccd$, and $cce$,
many elementary diagrams at the four-, five-, and higher-body level contributing
to $g_{ij}$ and $g_{ijk}$ can be evaluated.  
These central elementary diagrams also dress the SOCs.

In matter calculations, the correlations of Eq.\eqref{eq:heal} are generated by solving a set of coupled Euler-Lagrange equations in different pair-spin and isospin channels for $S=0,1$ and $T=0,1$.  
For pure neutron matter, only $T=1$ channels are needed, leaving a single-channel equation for $S=0$, producing a singlet correlation, and a triple-channel equation for $S=1$, which produces triplet, tensor, and spin-orbit correlations.  
The singlet and triplet correlations are then projected into central and $\sigma_{ij}$ combinations.  
Three (increasing) healing distances are used: $d_s$ for the singlet correlation, $d_p$ for the triplet and spin-orbit, and $d_t$ for the tensor.

Additional variational parameters are the quenching factors $\alpha_p$ whose introduction simulates modifications of the two--body potentials entering in the Euler--Lagrange differential equations arising from the screening induced by the presence of the nuclear medium 
\begin{equation}
v_{ij}=\sum_{p=1}^8 \alpha_p v^{p}(r_{ij})O^{p}_{ij}\, ,
\end{equation}
whereas the full potential is used when computing the energy expectation value.
In practice we use just two such parameters: $\alpha_{p=1} = 1$ and $\alpha_{p=2,8} = \alpha$.  
In addition, the resulting correlation functions $f^p$ may be rescaled according to  
\begin{equation}
F_{ij}=\sum_{p=1}^8 \beta_p f^{p}(r_{ij})O^{p}_{ij}\; ,
\end{equation}
with $\beta_{p=1} = 1$, $\beta_{p=2,4+7,8} = \beta_\sigma$, and $\beta_{p=5,6} = \beta_t$.  
These $\beta_p$ are not usually invoked when only two-body forces are considered, but they can significantly lower the variational energy when three-body forces are included.  
For the present work, the variational parameters are the three healing distances $d_c$, $d_p$, and $d_t$, one quenching factor $\alpha$, and two rescaling factors $\beta_\sigma$ and $\beta_t$.
These are varied at each density with a simplex search routine to minimize the energy.

The cluster diagrams for the three-body force are illustrated in Ref.~\cite{Carlson:1983kq}; Fig.2 shows the diagrams contributing to the $V^\Delta$ and $V^{2\pi}$ terms of Eqs.\eqref{eq:e51_delta}, and \eqref{eq:e51_2pi}, while Fig.3 shows diagrams contributing to the $c_E$ term of $V^{CT}$ of Eq.\eqref{eq:e51_ct}.  
Only the $c_D$ term of $V^{CT}$ requires some new work, but this is a straightforward generalization of the methods described in Ref.~\cite{Carlson:1983kq}; see also~\cite{Lovato:2011ij}.  
In the present work, the three-body force diagrams are also dressed with all appropriate central three-point superbonds $S_{xyz}$. 

One measure of the convergence of the FHNC/SOC integral equations is that the volume integral of the correlation hole from the central part of the two-body distribution function $g_{ij}$ (which has operator components like the $F_{ij}$ of Eq.~\eqref{eq:Foperator_v8prime}) should be unity.
To help guarantee that the variational parameters entering the FHNC/SOC correlations are well behaved -- and to ensure that in a given region of the parameter space the cluster expansion is converged -- we minimize the energy plus a constant times the deviation of the volume integral from unity:
\begin{equation}
 E + C \left\{ 1 + \rho\int d^{3}r \ [ g^c(r)-1 ] \right\}^2 \ , \nonumber
\end{equation}
as discussed in Ref.\cite{WFF88}.  
A value of $C=1000$ MeV is sufficient to limit the violation of this sum rule to 1\% or less at normal density, and 3\% or less at twice normal density for all the potentials considered here.  
In symmetric nuclear matter there is a related condition that the integral of the isospin component $g^\tau(r)$
\begin{equation}
 1 + \frac{1}{3}\rho\int d^{3}r \ g^\tau(r) \ , \nonumber
\end{equation}
should vanish to guarantee equal numbers of protons and neutrons.  
The corresponding integral with the spin component $g^\sigma(r)$ is not constrained when tensor forces are present.  
Instead, its deviation from unity provides a measure of the strength of the tensor (or spin-space) correlations in the system.

\subsection{AFDMC}
The AFDMC method~\cite{Schmidt:1999lik} projects out the ground-state of the system $|\Psi_0\rangle$ evolving a starting trial wave function $|\Psi_T\rangle$ in imaginary time $\tau$ as
\begin{equation}
|\Psi_0\rangle = \lim_{\tau \to \infty} |\Psi(\tau)\rangle = \lim_{\tau \to \infty} e^{-(H-E_T)\tau} |\Psi_T\rangle\, ,
\end{equation}
where $E_T$ is an estimate of the true ground-state energy $E_0$. The imaginary-time propagator $e^{-(H-E_T)\tau}$ is broken down and evaluated stochastically in $N$ small time steps $\delta \tau$, with $\tau= N \delta\tau$. At each step, the generalized coordinates $X^\prime$ are sampled from the previous ones according to the short-time propagator  
\begin{equation}
G(X^\prime,X,\delta\tau) = \frac{\Psi_I(X^\prime)}{\Psi_I(X)} \langle X^\prime| e^{-(H-E_0)\delta\tau}|  X \rangle
\end{equation}
where $\Psi_I(X^\prime)$ is the importance-sampling function. Similar to Refs.~\cite{Zhang:1996us,Zhang:2003zzk}, we mitigate the fermion-sign problem by first performing a constrained-path diffusion Monte Carlo propagation (DMC-CP), in which we take $\Psi_I(X) \equiv \Psi_T(X)$ and impose ${\rm Re}[\Psi_T(X^\prime)/\Psi_T(X)] > 0$. 
The solution obtained from the constrained propagation is not a rigorous upper-bound to $E_0$~\cite{Wiringa:2000gb}. To remove this bias, the configurations obtained from a DMC-CP propagation are further evolved using the positive-definite importance sampling function~\cite{Pederiva:2004iz,Lonardoni:2018nob,Piarulli:2019pfq}
\begin{align}
\Psi_I(X) = \sqrt{ \rm Re[\Psi_T(X)]^2 + \rm Im [\Psi_T(X)]^2}\,,
\end{align}
During this unconstrained diffusion (DMC-UC), the asymptotic value of the energy is determined by fitting its imaginary-time behavior with a single-exponential function~\cite{Pudliner:1997ck}. 

Applying standard diffusion Monte Carlo techniques to the nuclear many-body problem is made particularly complicated by the spin-isospin dependence of the nuclear forces. The AFDMC keeps the computational cost polynomial in the number of nucleons $A$ by representing the spin-isospin degrees of freedom in terms of outer products of single-particle states. To preserve this representation, during the imaginary-time propagation Hubbard-Stratonovich transformations are employed to linearize  the quadratic spin-isospin operators entering the nuclear potentials. When computing the AV18+UIX and NV2+3 Hamiltonians, the propagation is made with a simplified Hamiltonian $H^\prime$, which includes a re-projected $v_8^\prime$ version of the full $v_{18}$ potential~\cite{wiringa02}. The small difference $\Delta v = v_{18} - v_8^\prime$ is estimated at first order in perturbation theory as
\begin{align}
\langle \Delta v \rangle &\simeq 2 \frac{\langle \Psi_T | \Delta v | \Psi(\tau)\rangle }{ \langle \Psi_T | \Psi (\tau) \rangle } - \frac{\langle \Psi_T | \Delta v | \Psi_T\rangle }{ \langle \Psi_T | \Psi_T \rangle }\, .
\end{align}

No approximations are necessary to treat the three-body force in AFDMC calculations of PNM. As discussed in detail in Ref.~\cite{Pederiva:2004iz} for the UIX potential, the relation $\{\sigma_i^\alpha,\sigma_i^\beta\}=\delta^{\alpha\beta}$ is used to reduce the anticommutator terms of Eqs.~\eqref{eq:e51_delta} and~\eqref{eq:e51_2pi} to a sum of two-body spin operators with a form and strength that depend upon the positions of three particles. For each quantum Monte Carlo configuration, the sum over the position of the third particle is carried out explicitly so that including these terms only involves changing the strength of the $N\!N$ potential's spin matrices. Similar strategies can be followed to handle the $c_1$ term in Eq.~\eqref{eq:e51_2pi}, which is equivalent to the S-wave component of the Tucson-Melbourne three-body potential~\cite{Coon:1980fa}, and the $c_D$ contribution~\cite{Lovato:2011ij}. Finally, in PNM the $c_E$ term of Eq.~\eqref{eq:e51_ct} is identical to the phenomenological repulsive scalar contribution of the UIX force --- again with different radial functions and coupling constant --- and its inclusion is trivial~\cite{Pederiva:2004iz}.

Within the AFDMC, infinite uniform neutron matter is typically simulated using a finite number of neutrons obeying periodic-box boundary conditions (PBC)~\cite{Pederiva:2004iz}. The trial wave function adopted in our calculations respects by construction these PBC and it is expressed as 
\begin{equation}
\Psi_T(X) = \langle X | \Psi_T\rangle = \langle X | \prod_{i<j} f_c(r_{ij}) |\Phi\rangle\, .
\end{equation}
The spin-isospin independent Jastrow function $f_c(r_{ij})$ is parametrized in terms of a cubic spline that has a smooth first derivative and continuous second derivative. The variational parameters to be optimized are the values of the spline at the grid points, plus the value of the first derivative at $r_{ij}=0$. In order for the PBC to be satisfied, we impose $f_c(r_{ij})=1$ and $f_c^\prime(r_{ij}) =0$ when $r_{ij}\geq L/2$. This Jastrow ansatz is more flexible than the one used in our previous work~\cite{Piarulli:2019pfq} and has proven to be able to capture highly clustered configurations of nucleons, as discussed in Ref.~\cite{Contessi:2017rww} for the $^{16}$O nucleus. 

In neutron-matter calculations, the mean-field part of the wave function is usually taken to be the plane-waves Slater determinant of Eq.~\eqref{def:phi}. In order to satisfy the PBC, the single-particle wave vectors take the discrete values
\begin{equation}
\mathbf{k}_i=\frac{2\pi}{L} \{ n_x,n_y,n_z\}\quad,\quad n_i=0,\pm 1,\pm 2,\dots\, ,
\end{equation}
$L$ being the size of the box. When simulating homogeneous and isotropic systems, calculations are performed with closed momentum shells. Since neutrons can be spin-up or spin-down, this corresponds to $A=2$, $14$, $38$, $54$, $66$, $114$, etc., particles in a box. In this work, we improve upon the plane-wave Slater determinant by including the spin-dependent backflow correlations~\cite{Brualla:2003gw} through the following replacement
\begin{equation}
e^{i \mathbf{k}_i \cdot \mathbf{r}_j} \to e^{i \mathbf{k}_i \cdot \mathbf{r}_j} \bigg[1 + \frac{1}{2} \sum_{k\neq j} f_b(r_{jk}) (\mathbf{r}_{jk} \times \mathbf{k}_i)\cdot \boldsymbol{\sigma}_j\bigg]\, .
\end{equation}
In our previous work, the spin-orbit Jastrow function $f^b(r_{ij})$ was determined minimizing the two-body cluster contribution to the energy per particle, similarly to what is done in the FHNC/SOC method. In that case, we observed relatively minor improvements in the energy per particle with respect to the simpler plane-wave Slater determinant. Here, we adopt a more flexible cubic spline parametrization, analogous to the one used for $f^c(r_{ij})$; the only difference with the latter is that to fulfill PBC we impose $f_b(r_{ij})=0$ and $ f_b^\prime(r_{ij})=0$ when $r_{ij}\geq L/2$. As shown in Table~\ref{tab:afdmc_1}, this ansatz lowers the VMC energies per particle by $1.3$ MeV and $2.0$ MeV per nucleon at saturation density, for the AV18+UIX and NV2+3-Ia$^*$ Hamiltonians, respectively. More importantly, the DMC-CP results are also noticeably improved, and they are closer to their DMC-UC values: $18.1(3)$ and $15.3(3)$ MeV for the AV18+UIX and the NV2+3-Ia$^*$ models, respectively. This behavior demonstrates that the flexible cubic-spline spin-dependent backflow correlations refine the phase structure of the variational wave function, making it closer to the one of the true ground-state of the system, thereby improving the accuracy of the constrained-path approximation. In contrast to the linearized spin-dependent correlations introduced in Ref.~\cite{Gandolfi:2013baa}, these spin-dependent backflow correlations add minor overhead to the computational cost of the method, which still scales as $A^3$. In addition, their inclusion does not violate the factorization theorem, and it is therefore better suited for treating translation-invariant uniform systems. 

\begin{center}
\begin{table}
\begin{tabular}{l| c c c c c}
\hline\hline
 & VMC$_{\rm PW}$ & VMC$_{\rm BF}$  & DMC-CP$_{\rm PW}$ & DMC-CP$_{\rm BF}$\\
\hline
AV18+UIX  &    23.13(2)  &  21.78(1) &   20.80(4)    & 19.32(3) \\
NV2+3-Ia$^*$  &  20.40(1) & 18.42(1) & 17.31(4)    & 16.32(2)\\
\hline\hline
\end{tabular}
\caption{Energies per particle of PNM at $\rho = 0.16$~fm$^{-3}$ obtained with the plane-wave (PW) and spin-dependent backflow (BF) Slater determinant using the AV18+UIX and NV2+3-Ia Hamiltonians simulating 66 neutrons with PBC.}
\label{tab:afdmc_1}
\end{table}
\end{center}

The DMC-CP simulations presented in this work are carried out employing $A=66$ neutrons in a box. This choice considerably reduces finite-size effects, since the kinetic energy of $66$ fermions approaches the thermodynamic limit very well~\cite{Gandolfi:2009fj}. In addition, finite-size effects due to the tail corrections of two- and three-body potentials are accounted for by summing the contributions given by neighboring cells to the simulation box~\cite{Sarsa:2003zu}.
When not otherwise specified, the AFDMC results presented in this work always refer to the unconstrained energies. The latter are obtained by adding to the DMC-CP values obtained with 66 neutrons the difference between DMC-UC and DMC-CP energies computed with 14 neutrons with PBC, applying tail corrections to the potential-energy contribution as discussed above. This procedure decreases the computational cost of the method, which still requires about $150$k core-hours for each value of the density for a given Hamiltonian. Its accuracy has been validated in Ref.~\cite{Piarulli:2019pfq} by performing DMC-UC simulations with 14 and 38 neutrons, which showed negligible differences. Finally, we note that the unconstrained energies are independent of the trial wave function of choice, although including backflow correlations significantly reduce the statistical noise of our results.

\section{Results}
\label{sec:results}
\begin{figure}[t]
\includegraphics[width=\linewidth]{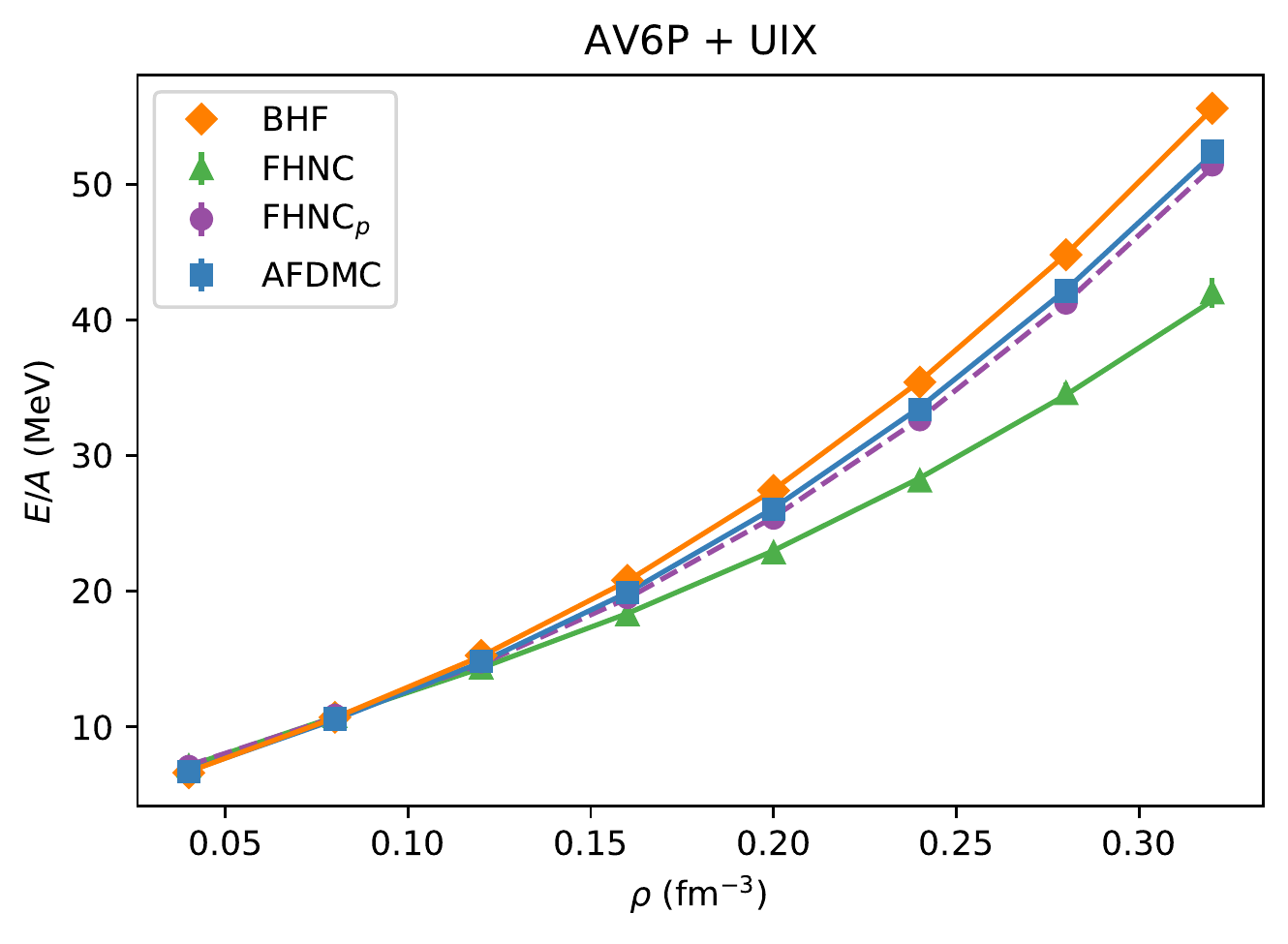}
\includegraphics[width=\linewidth]{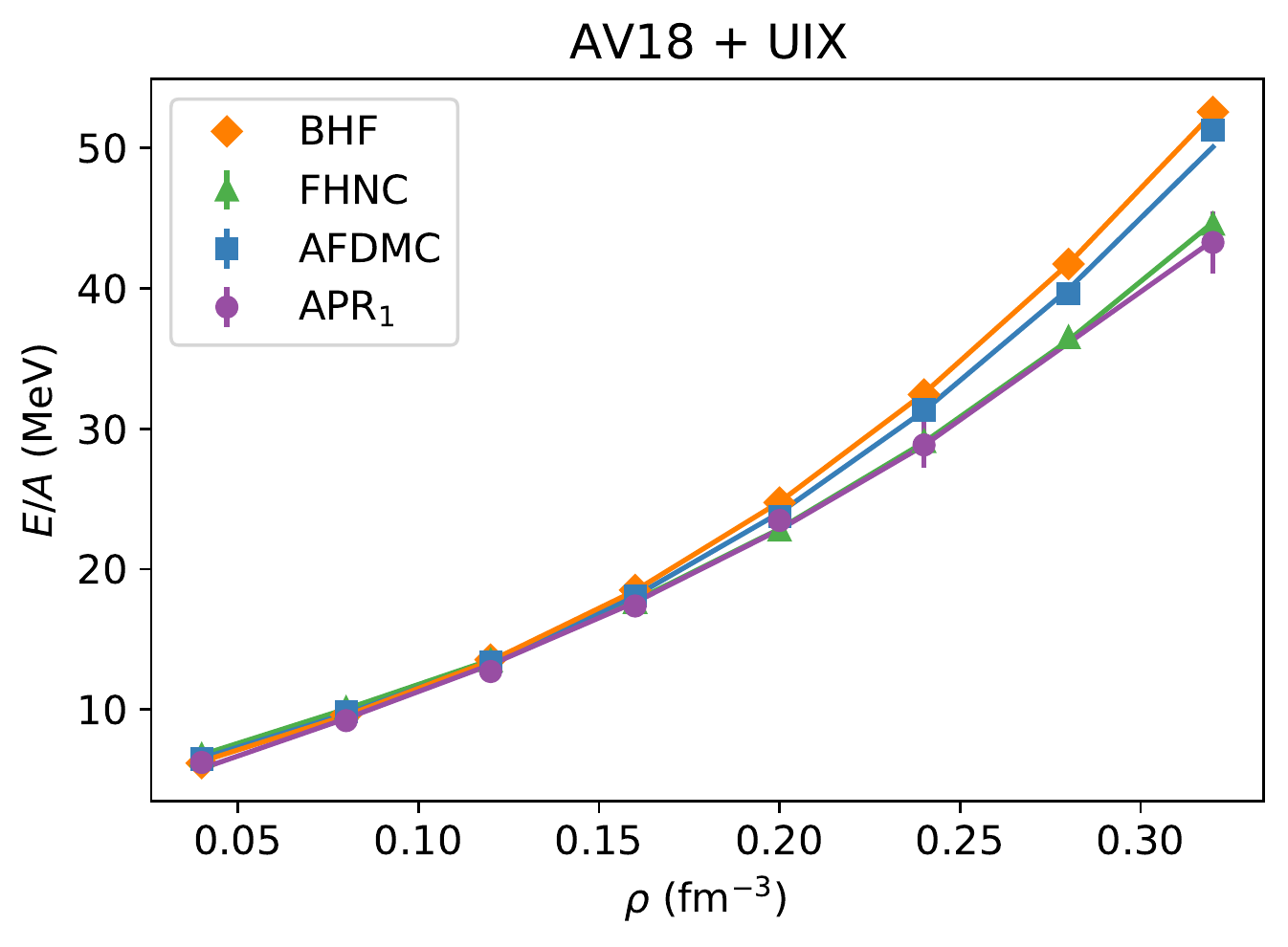}
\caption{Neutron-matter EOS as obtained from the AV6P+UIX (upper panel) and AV18+UIX (lower panel) Hamiltonians. ~\label{fig:EOS_AV}}
\end{figure}

In Fig.~\ref{fig:EOS_AV} we display the EOS of PNM using the semi-realistic, phenomenological AV6P+UIX (upper panel) and highly-realistic AV18+UIX (lower panel) Hamiltonians. The BHF, FHNC/SOC, and AFDMC many-body methods yield almost identical energies per particle up to saturation density for both interaction models. At $\rho>\rho_0$, the FHNC/SOC results lie below the BHF and AFDMC ones, consistent with what was observed in Ref.~\cite{Piarulli:2019pfq} for the $N\!N$ interaction only.

For the specific case of the AV6P+UIX Hamiltonian, the discrepancies between the FHNC/SOC and the other two many-body methods are larger than with the $N\!N$ potential alone.  As the density increases, the optimized correlations change character, with large values of $\beta_t$ that significantly enhance the normally small tensor correlations in neutron matter from $N\!N$ forces. This can change the expectation value of the two-pion exchange part of UIX from weak repulsion to strong attraction.  For AV6P alone, the optimized $\beta_t$ vary only slightly from unity with no significant lowering of the energy.  If UIX is added in perturbation, i.e., without re-optimizing the variational parameters, the result, shown by the dashed line labeled FHNC$_p$ in the upper panel of Fig.~\ref{fig:EOS_AV}, is very close to the AFDMC result.

This kind of behavior of the FHNC/SOC energies has been observed before~\cite{Akmal:1997ft} where it was identified as a neutral pion condensate, i.e., a tendency toward spin-space order in nucleon matter, and it is a noticeable feature of the WFF and APR EOS~\cite{WFF88,Akmal:1998cf}.  The FHNC/SOC calculations for AV18+UIX shown in the lower panel of Fig.~\ref{fig:EOS_AV} also have this enhanced tensor correlation, but the lowering of the energy relative to the BHF and AFDMC calculations is less dramatic than for AV6P+UIX.  This kind of ordered solution in FHNC/SOC calculations depends on the Hamiltonian, and is not found in most of the Norfolk potentials -- only model NV2+3-Ib* seems to exhibit this behavior.  Whether this kind of solution is a result of approximations in the treatment of the $3N$ potential contribution~\cite{Carlson:1983kq}, or it is actually discovering a phase that the BHF and AFDMC calculations miss, is unknown at this time.

For reference, in Fig.~\ref{fig:EOS_AV}, we show the APR result of Ref.~\cite{Akmal:1998cf}, which has also been obtained solving the ground-state of AV18+UIX with the FHNC/SOC method. As in Ref.~\cite{Sabatucci:2020xwt}, we denote this EOS ``APR$_1$'' to differentiate it from the one that includes relativistic corrections arising from the boost of the $N\!N$ potential, dubbed ``APR$_2$''. The differences between APR$_1$ and our own FHNC/SOC results are minimal and mainly due to our improved treatment of elementary diagrams and some differences in the handling of spin-orbit correlations.  

On the other hand, the AFDMC and BHF energies per particle remain very close up to $\rho = 0.32$ fm$^{-3}$ --- the maximum difference being $3.2$ MeV and $2.1$ MeV for AV6P+UIX and AV18+UIX, respectively. 
 The AFDMC and BHF EOS obtained with the AV18+UIX interaction are remarkably similar to APR$_1$ up to $\rho_0$, thereby corroborating its accuracy, but become stiffer at higher densities. This behavior is consistent with a recent Bayesian analysis of the of masses, radii, and tidal deformabalities measured by the NICER satellite and the LIGO/Virgo collaboration that favors a stiffer EOS than APR$_2$~\cite{Maselli:2020uol}. However, before definitive conclusions can be drawn, relativistic corrections must be included in both AFDMC and BHF calculations. It has also to be noted that cluster variational Monte Carlo calculations indicate that the AV18+UIX Hamiltonian underbinds $^{16}$O and $^{40}$Ca nuclei~\cite{Lonardoni:2017egu}. A less repulsive version of the UIX force is therefore required to reproduce the ground-state energies of these nuclei, which will likely soften the EOS of PNM. 

The differences between the AV6P+UIX and AV18+UIX EOS are much smaller that when the two-body forces alone are included~\cite{Baldo:2012nh,Piarulli:2019pfq}. This behavior may be ascribed to the phenomenological repulsive term of the UIX potential that prevents nucleons from getting close to each other, thereby reducing the relevance of accurately fitting high partial waves in $N\!N$ scattering. Hence, as argued by the Authors of Refs.~\cite{Benhar:2017oli}, the AV6P+UIX Hamiltonian can be safely employed to make predictions of neutron-matter properties, including finite-temperature ones~\cite{Camelio:2017nka}, avoiding the technical complications associated with spin-orbit and quadratic spin-orbit operators.  

The curves in the plot correspond to a polynomial fit for the density dependence of the energy per particle. Its functional form contains a term proportional to the kinetic energy of a free Fermi gas plus two contributions that are inspired by a cluster expansion of the energy expectation value, truncated at the three- body level
\begin{equation}
\frac{E(\rho)}{A} = a_{2/3} \left(\frac{\rho}{\rho_0}\right)^{2/3} + a_{1} \left(\frac{\rho}{\rho_0}\right) + a_{2} \left(\frac{\rho}{\rho_0}\right)^2\, .
\label{eq:pol_fit}
\end{equation}
The fit parameters that best reproduce the AFDMC calculations are listed in Table~\ref{tab:av_fit} with their estimated errors. The covariance matrix of the fit has also been computed and it is available upon request. 
\begin{center}
\begin{table}[t]
\begin{tabular}{l| c c c}
\hline\hline
 & $a_{2/3}$ & $a_1$ & $a_2$ \\
\hline
AV6P+UIX  &    $27.2 \pm 0.4$ &  $-19.1\pm 0.8$ & $11.80 \pm 0.04$    \\
AV18+UIX  &    $30.5 \pm 1.1$  &  $-25.7\pm 2.4$ &   $13.24\pm0.23$  \\
\hline\hline
\end{tabular}
\caption{Best-fit parameters from Eq.~\eqref{eq:pol_fit} for the AFDMC energy per particle obtained from the AV6P+UIX and AV18+UIX Hamiltonians.}
\label{tab:av_fit}
\end{table}
\end{center}
\begin{figure}[b]
\includegraphics[width=\linewidth]{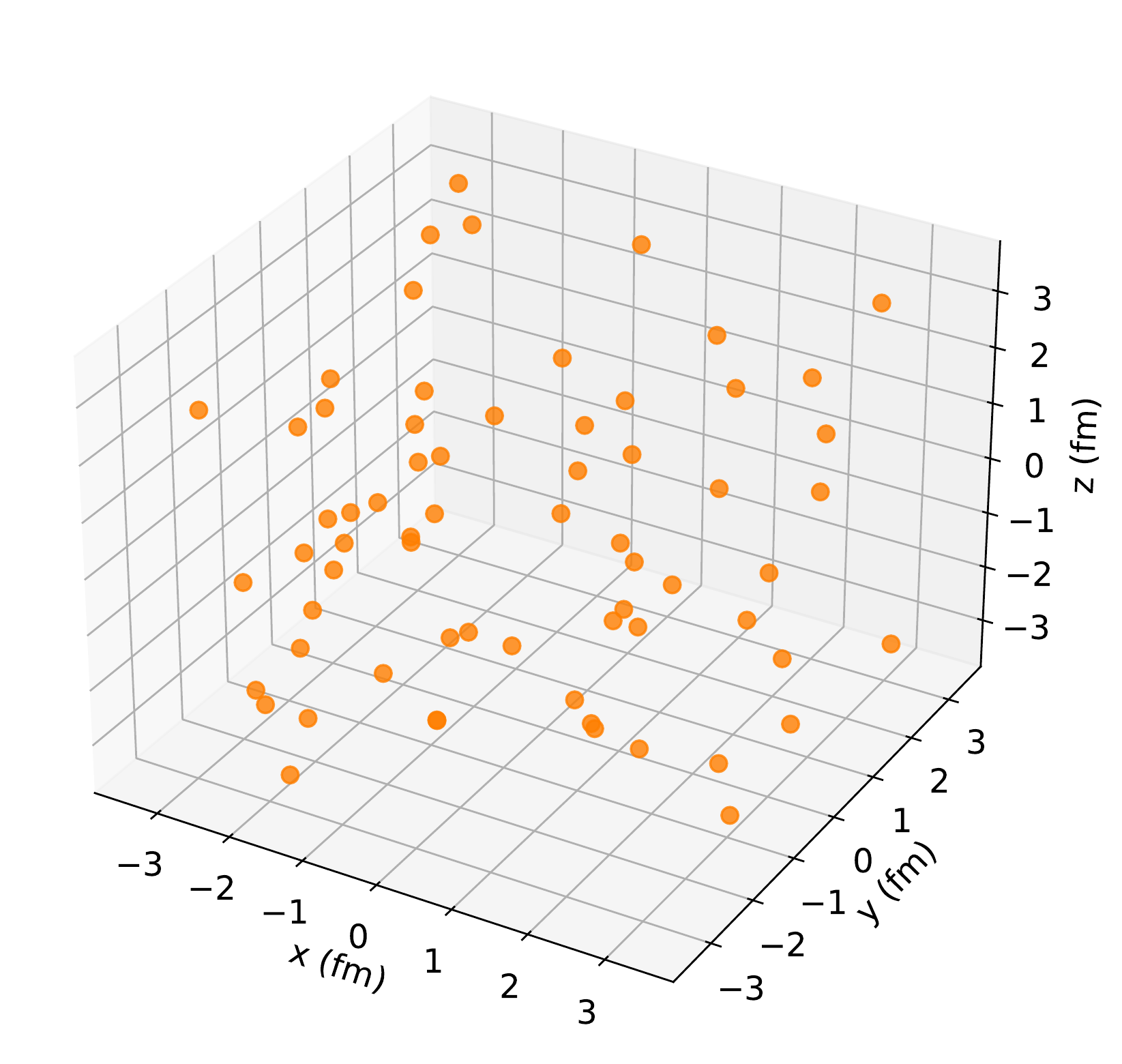}
\includegraphics[width=\linewidth]{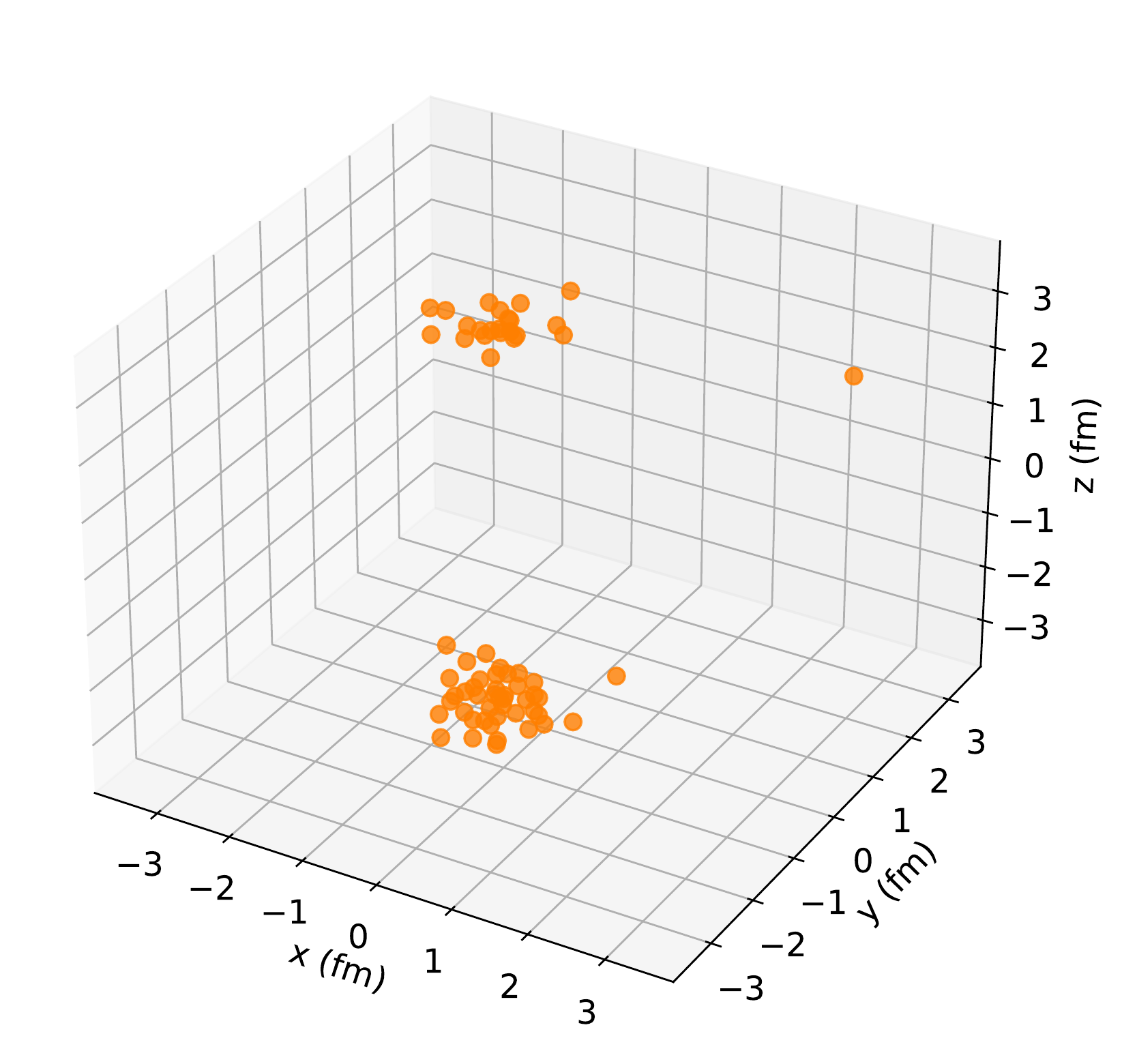}
\caption{Single snapshot of a Metropolis random walk for VMC calculations. The variational wave functions are optimized with the NV2-Ia two-body force alone (upper panel) and including the three-body force NV2+3-Ia (lower panel) which leads to the formation of neutron droplets.~\label{fig:collapse}}
\end{figure}

The first generation of Norfolk $N\!N$ plus $3N$ Hamiltonians, fitted on the trinucleon ground-state energies and $nd$ doublet scattering length, are characterized by relatively large and negative values of $c_E$, listed in Table~\ref{tab:cdce}. When used as inputs in the AFDMC, all the NV2+3-Ia/b and NV2+3-IIa/b Hamiltonians yield to the ``collapse'' of PNM, whose energy per particles became large --- of the order of several GeV per particle --- and negative already at saturation density. Thanks to the flexibility of our variational ansatz, based on cubic-spline correlations, the collapse is clearly visible already at the variational level. On the other hand, using correlation functions determined minimizing the two-body cluster contribution to the energy per particle, as done in our previous work~\cite{Piarulli:2019pfq}, prevents the collapse from happening at the VMC level. In this latter case, PNM becomes deeply bound already after a few time steps in the imaginary-time diffusion. 

The collapse is associated with the formation of ``droplets'' of closely packed neutrons, ultimately caused by the attractive nature of the $c_E$ term in the $3N$ force. Its strength grows with the third power of the number of particles in a droplet, and overcomes the repulsive kinetic-energy contribution. To better illustrate this behavior, in Fig.~\ref{fig:collapse} we display the positions of $66$ neutrons with PBC obtained from a single Metropolis step of a variational Monte Carlo calculation for model NV2+3-Ia. In the upper panel, the $3N$ force is turned off and the neutrons are distributed uniformly in the box. When the $3N$ is included in the Hamiltonian, the variational wave function changes dramatically, making the neutrons form closely-packed droplets. Note that the average density of the system is unchanged, as the droplets move across the box --- and in fact they can enter nearby boxes so that periodicity is enforced.

Requiring the energy per particle of PNM to be positive at $\rho=\rho_0$ yields lower bounds on $c_E$. We find that these limits are fairly insensitive to the value of $c_D$ --- whose impact in PNM is modest --- and, more surprisingly, to the specific $N\!N$ interaction of choice. In fact, taking $c_E \gtrsim -0.1$ is sufficient to avoid the collapse, for all the NV2+3-Ia/b and NV2+3-IIa/b models. These limits are conservative for primarily two reasons. First, we have obtained them by simulating $66$ neutrons with PBC. At fixed density, the expectation value of the $3N$ force grows a factor $\sim N$ faster than the $N\!N$ potential and a factor $\sim N^2$ faster than the kinetic energy, where $N$ is the number of neutrons in the box. Hence, putting more neutrons in the box will likely increase the relative importance of the $3N$ interaction, bringing the lower limits on $c_E$ closer to zero --- see Ref.~\cite{Gridnev:2013twa} for a mathematical discussion on this point. Second, here we are only imposing positive energies per particle, neglecting constraints coming from astrophysical observations, such as the maximum mass of the star or its tidal deformability, which will probably require stiffer EOS, and hence more stringent limits on $c_E$.

The FHNC/SOC calculations for NV2+3-Ia/b and -IIa/b also find these models are generally not suitable for neutron star EOS.  Finite energies are found at each density, but models Ia and IIa, with their large negative $c_E$ terms, have energy maxima near saturation density $\rho_0$ and become less repulsive or even bound at higher densities, indicative of collapse.  Energies for models Ib and IIb continue to increase slowly up to $2\rho_0$, but NV2+3-Ib is already less repulsive than NV2-Ib alone.  Model IIb shows the greatest stability, consistent with having the least negative $c_E$ term, but the energy appears to be near a maximum at $2\rho_0$.  As mentioned above, the FHNC/SOC energies for these models show relatively little sensitivity to the variational $\beta_t$ parameter, i.e., no evidence for a neutral pion condensate.

\begin{figure}[h]
\includegraphics[width=0.99\linewidth]{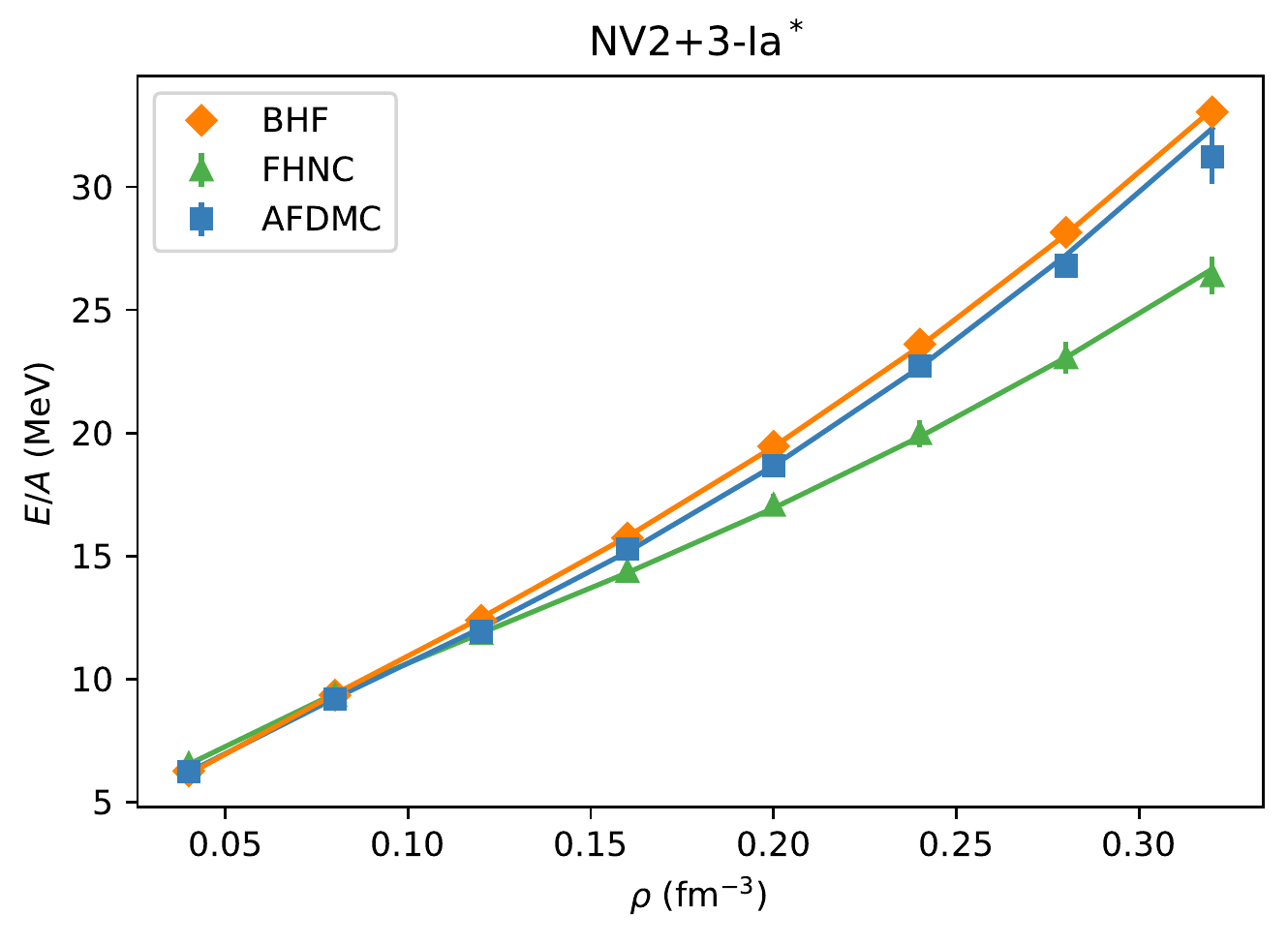}
\includegraphics[width=0.99\linewidth]{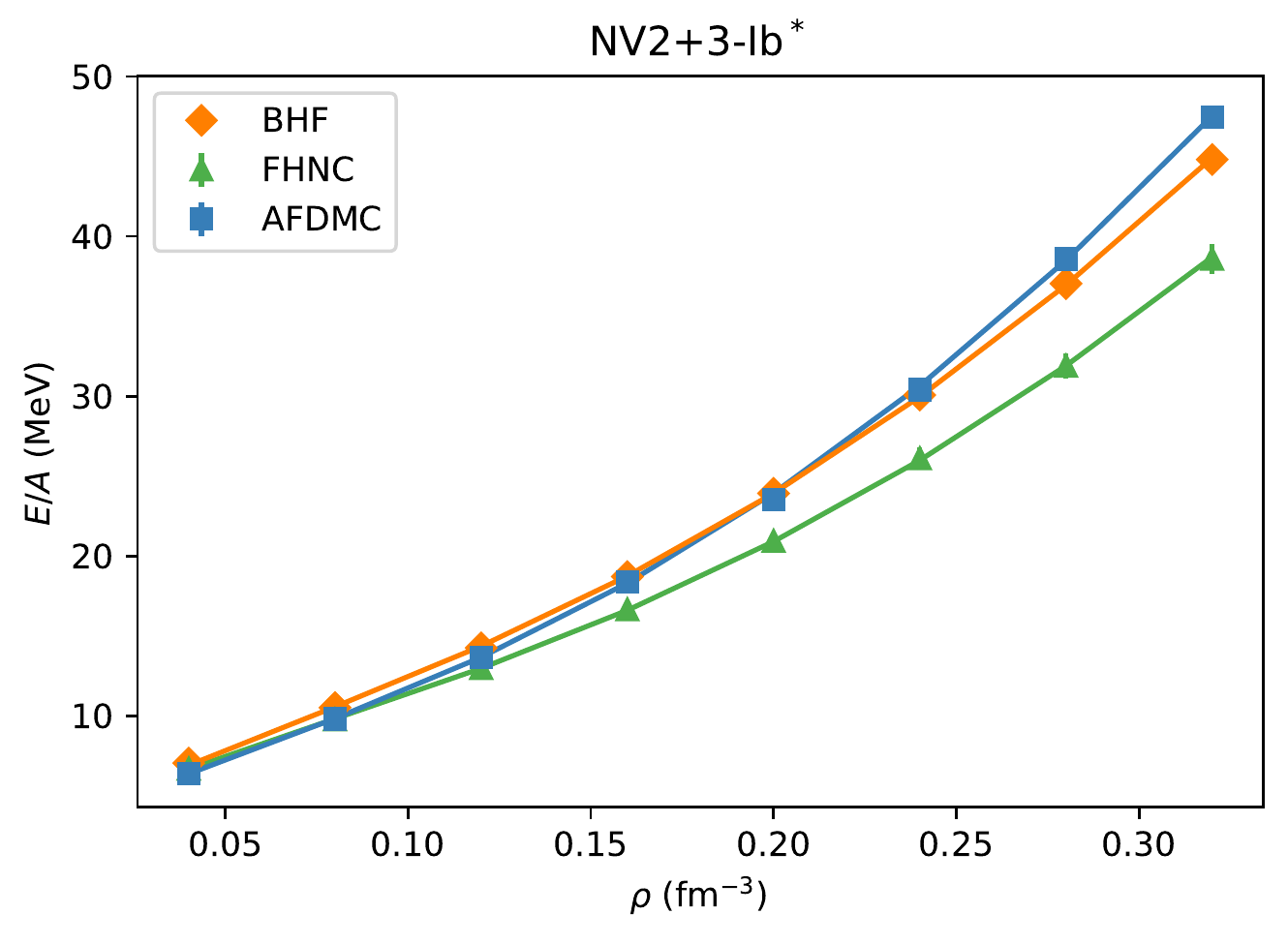}
\includegraphics[width=0.99\linewidth]{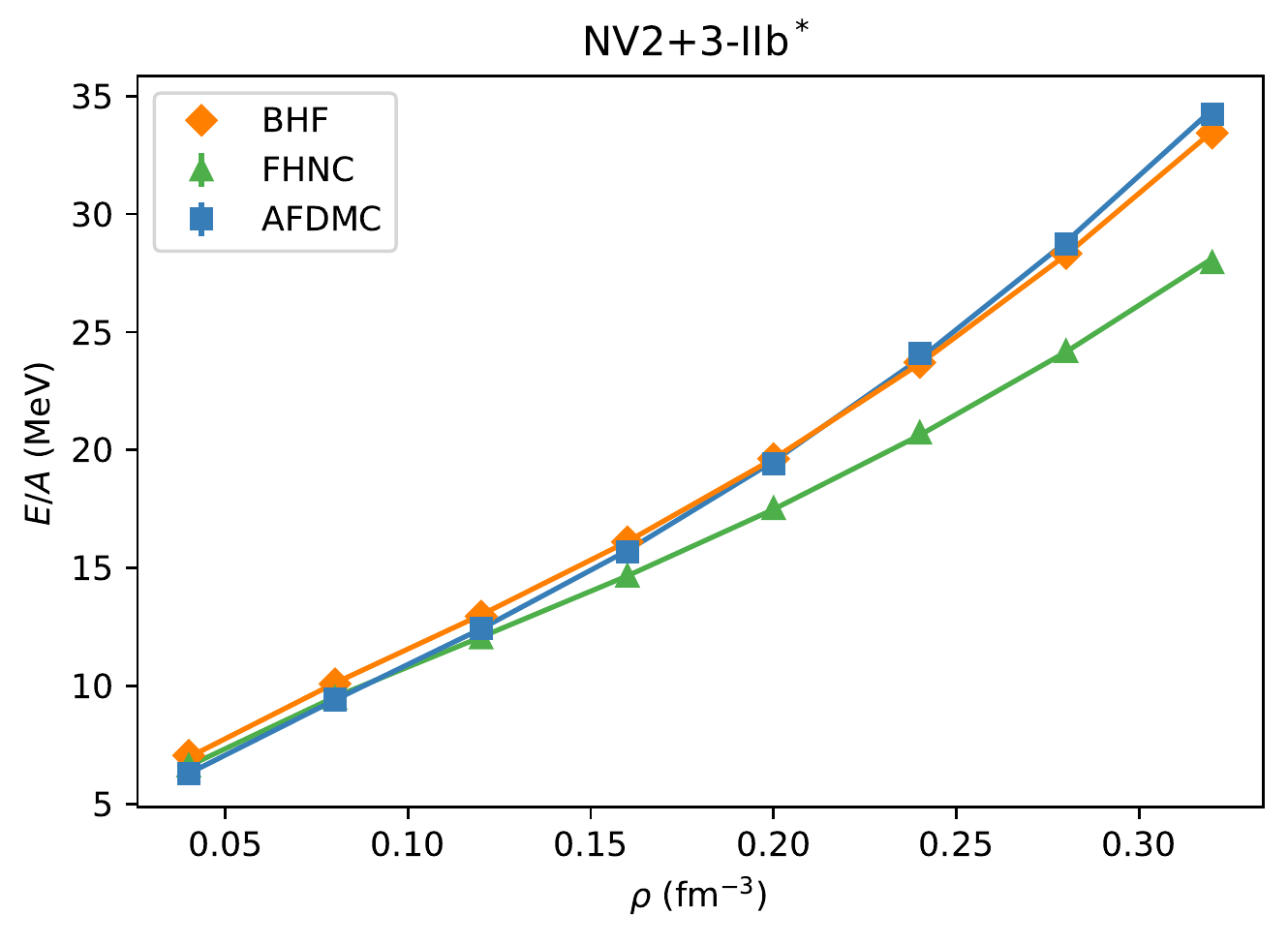}
\caption{Pure neutron-matter EOS as obtained from the NV2+3-1a$^*$ (upper panel) and NV2+3-1b$^*$ (middle panel), and NV2+3-2b$^*$ Hamiltonians. ~\label{fig:EOS_NV}}
\end{figure}

The NV2+3-Ia*/b* and NV2+3-IIa*/b* Hamiltonians are characterized by smaller values of $c_E$ than NV2+3-Ia/b and NV2+3-IIa/b. As a consequence, among the models fitted to also reproduce tritium $\beta$ decay, only NV2+3-IIa* causes PNM to collapse at $\rho=\rho_0$. In the latter case however, the uncertainty in $c_E$, found by propagating the experimental error of $\beta$ decay rate, is about $0.1$. Therefore, it may be possible to find a value for $c_E$ for this model that does not yield collapsing PNM while still providing a $\beta$-decay rate compatible with the experimental value, at least within three-sigma. 

The EOS obtained with all the other models using the BHF, FHNC/SOC, and the AFDMC methods are displayed in Fig.~\ref{fig:EOS_NV}. The solid curves correspond to the polynomial fit of Eq.~\eqref{eq:pol_fit}, whose best parameters for the AFDMC method are listed in Table ~\ref{tab:nv_fit}. Similarly to the AV6P+UIX and AV18+UIX cases, the BHF and AFDMC energies are remarkably close up to twice saturation density --- the maximum difference remaining within $2.7$ MeV per particle. On the other hand, the EOS computed within the FHNC/SOC method are softer, especially in the high-density region. It is however remarkable that the three many-body methods differ at most by $1.9$ MeV per particle for $\rho \leq \rho_0$. Extending the comparison to the high-density region, the discrepancies among the many-body methods remain below $5.9$ MeV per particle, and hence significantly smaller than the $16.2$ MeV difference between the AFDMC results obtained with the NV2+3-Ia* and NV2+3-Ib* Hamitlonians at $\rho= 2\rho_0$. Hence, the theoretical uncertainty associated with modeling nuclear dynamics is more relevant than the one pertaining to the many-body methods --- even excluding from this comparison the Hamiltonians that yield a deeply-bound EOS of PNM. 

\begin{center}
\begin{table}[t]
\begin{tabular}{l| c c c}
\hline\hline
 & $a_{2/3}$ & $a_1$ & $a_2$ \\
\hline
NV2+3-Ia*  &    $24.23 \pm 0.44$ &  $-15.09\pm 0.99$ & $6.02 \pm 0.13$    \\
NV2+3-Ib*  &    $26.17 \pm 0.18$  &  $-18.71\pm 0.40$ &   $10.85 \pm 0.05$  \\
NV2+3-IIb*  &    $24.35 \pm 0.46$  &  $-15.11\pm 0.90$ &   $6.49 \pm 0.06$  \\
\hline\hline
\end{tabular}
\caption{Best-fit parameters from Eq.~\eqref{eq:pol_fit} for the AFDMC energy per particle obtained from the NV2+3-Ia*/b* and NV2+3-IIb* Hamiltonians.}
\label{tab:nv_fit}
\end{table}
\end{center}

\section{Conclusions}
\label{sec:conclusions}
Using as input the phenomenological AV6P+UIX and AV18+UIX Hamiltonians, we observe an excellent agreement between the AFDMC, BHF, and FHNC/SOC many-body approaches up to saturation density, while in the region $\rho>\rho_0$ the FHNC/SOC yields softer EOS than the other two methods, consistent with our $N\!N$ potential results~\cite{Piarulli:2019pfq}. Our updated version of the FHNC/SOC method, which includes classes of elementary diagrams that were previously neglected, is in excellent agreement with the APR$_1$ EOS. However, our AFDMC and BHF results indicate that the AV18+UIX Hamiltonian produces an EOS which is even stiffer than APR$_1$. Before comparisons with astrophysical observations can be made, relativistic boost corrections, which are known to soften the EOS in the high-density regime, must be included in our calculations. Finally, we observe that the repulsive component of the UIX potential seems to reduce the importance of high partial waves in the $N\!N$ scattering. As a consequence, the EOS obtained with the full AV6P+UIX and AV18+UIX Hamiltonians are much closer to each other than those obtained with then $N\!N$ force alone, independent of the many-body method of choice. This behavior 
suggests the utility of using the AV6P+UIX Hamiltonian for studying PNM properties, including finite-temperature ones, without the technical complications that spin-orbit and quadratic spin-orbit operators involve. 

The AFDMC calculations indicate that the first generation of Norfolk potentials yields deeply bound PNM already at saturation density. This behavior, which is even more extreme than already found in Refs.~\cite{Lovato:2011ij,Lynn:2015jua,Lonardoni:2019ypg}, is driven by the attractive nature of the $c_E$ component of the $3N$ force, when the operator $\tau_{ij}$ is chosen among the Fierz-equivalent ones. This term provides a non-negligible contribution to the energy per particle of PNM when regulators that are local in coordinate space are used, and only vanishes in the infinite regulator limit. The flexibility of the AFDMC trial wave function introduced in this work makes it possible to observe the collapse of PNM already at the variational level. On the other hand, the BHF and the FHNC/SOC methods do not provide such stringent indications of a collapse. The former relies on a density dependent formulation of the $3N$ potential, similar to the normal ordering approximation used in other many-body methods~\cite{Hagen:2007ew,Roth:2011vt}. The latter instead is based on correlation functions that are largely determined minimizing the two-body cluster contribution to the energy per particle. These approximations are reliable only when the $3N$ force is much weaker than the $N\!N$ one. However, this is not the case for the attractive $c_E$ contribution, as it induces significant changes in the ground-state wave function, leading to the formation of closely packed droplets of neutrons. 

Our analysis reiterates the limitations of fitting the LECs of the $3N$ force to reproduce highly-correlated observables. However, since the NV2+3-Ia model accurately reproduces many ground- and excited state properties of light nuclei~\cite{Piarulli:2017dwd}, there are no reason to expect dramatically different values of $c_D$ and $c_E$ if these latter quantities were included in the fit. Analogously to the AV18+IL7 Hamiltonian, an accurate description of light-nuclear spectra does not automatically translate into reliable predictions for the EOS of PNM. The reason behind this common behavior is likely that $A \leq 12$ nuclei are either almost isospin-symmetric or too light to constrain the behavior of $c_E$ in dense neutron matter. The above problems are mitigated by determining $c_D$ and $c_E$ to simultaneously reproduce the tri-nucleon binding energies and $\beta$ decay rates. Once again our three many-body methods of choice, and in particular the AFDMC and BHF, provide similar EOS of PNM for the NV2+3-Ia*/b* and NV2+3-IIb* models up to twice saturation density --- the FHNC/SOC results are softer, consistent with what found using phenomenological Hamiltonians. However, as shown by AFDMC calculations with the for NV2+3-IIa* interactions, the problem of deeply bound PNM is not fully solved by including the $\beta$ decay in the fit.

The benchmark calculations between the BHF, FHNC/SOC, and AFDMC methods indicate that the main source of theoretical uncertainty in the EOS of PNM comes from modeling nuclear dynamics rather than the many-body method used to solve the Schr\"odinger equation. The highly-realistic Hamiltonian AV18+UIX provides an EOS that satisfies astrophysical constraints, but it underbinds light- and medium-mass nuclei~\cite{Lonardoni:2017egu}. In addition, being phenomenological, it does not come with a clear-cut way to estimate its theoretical uncertainty. On the other hand, the Norfolk local potentials derived within $\chi$EFT provide an accurate description of light-nuclei spectra, but their PNM predictions suffer from significant regulator artifacts, hampering their predictive power. To remedy this shortcoming, as in Ref.~\cite{Lynn:2015jua}, other choices than $\tau_{ij}$ for the $c_E$ term should be considered, such as the identity operator and the projector on triplets with $S=1/2$ and $T=1/2$ --- the only ones contributing in the infinite-regulator limit. In addition, as argued in the illuminating Ref.~\cite{Huth:2017wzw}, including subleading terms in the chiral expansion should reduce these regulator artifacts. 

AFDMC calculations of oxygen isotopes will help shed some light on this problem, since their binding energies and radii should be more sensitive to the details of the three-nucleon interaction, including its isospin dependence, than lighter systems~\cite{Otsuka:2009cs,Lapoux:2016exf}. Complementary finite systems that are ideally suited to test the behavior of three-neutron interactions are the ones comprised of neutrons confined by an external potential, or ``neutron drops''~\cite{Pudliner:1995jn,Gandolfi:2010za,Maris:2013rgq,Potter:2014dwa,Zhao:2016ujh,Shen:2018euq}. Based on the findings of the present work, once enough neutrons are added to the system so that the $3N$ force becomes dominant, it is plausible that the attractive $c_E$ contribution makes neutron drops self bound, even with no external confining potential. Since no such systems are observed in nature, this fact would impose additional constraints on the $3N$ interaction.

\section{Acknowledgements}
We thank Omar Benhar, Richard J. Furnstahl, Stefano Gandolfi, Francesco Marino, Francesco Pederiva, Daniel Phillips, and Ragnar Stroberg for useful discussions. This research is supported by the U.S. Department of Energy, Office of Science, Office of Nuclear Physics, under contract DE-AC02-06CH11357 and the NUCLEI SciDAC program (A. L. and R. B. W.), the 2020 DOE Early Career Award number ANL PRJ1008597 (A. L.), Argonne LDRD awards, the 2021 Early Career Award number DE-SC0022002 (M. P.), and the FRIB Theory Alliance award DE-SC0013617 (M. P.). A.L acknowledges funding from the INFN grant INNN3, and from the European Union Horizon 2020 research and innovation programme under grant agreement No 824093.
The many-body calculations were performed on the parallel computers of the Laboratory Computing Resource Center, Argonne National Laboratory, the computers of the Argonne Leadership Computing Facility via the 2019/2020 ALCC grant ``Low Energy Neutrino-Nucleus interactions'' for the project NNInteractions, and the 2020/2021 ALCC grant ``Chiral Nuclear Interactions from Nuclei to Nucleonic Matter'' for the project ChiralNuc, and through a CINECA-INFN  agreement, providing access to resources on MARCONI at CINECA.

\bibliography{biblio}

\end{document}